\documentclass[aps,prb,twocolumn,showpacs,superscriptaddress, longbibliography]{revtex4-1}

\usepackage{xcolor, graphicx}
\usepackage{amssymb,amsmath, amsfonts, bm}
\usepackage{newtxtext,newtxmath}

\usepackage{dcolumn}
\usepackage[none]{hyphenat} 
\usepackage{tabularx}

\usepackage{newtxtext,newtxmath}

\usepackage{verbatim}  

\begin{document}

\title{Multi-type Dirac fermions protected by orthogonal glide symmetries in a noncentrosymmetric system}

\author{Yunyouyou Xia}
\affiliation{School of Physical Science and Technology, ShanghaiTech University, Shanghai 200031, China}
\affiliation{Shanghai Institute of Optics and Fine Mechanics, Chinese Academy of Sciences, Shanghai 201800, China}
\affiliation{University of Chinese Academy of Sciences, Beijing 100049, China}
\author{Xiaochan Cai}
\affiliation{School of Physical Science and Technology, ShanghaiTech University, Shanghai 200031, China}
\author{Gang Li}
\email{ligang@shanghaitech.edu.cn}
\affiliation{School of Physical Science and Technology, ShanghaiTech University, Shanghai 200031, China}
\affiliation{\mbox{ShanghaiTech Laboratory for Topological Physics, ShanghaiTech University, Shanghai 200031, China}}

\begin{abstract}
Compared to inversion-symmetric systems, emerging bulk Dirac point (DP) in noncentrosymmetric systems is much harder and its symmetry protection mechanism is poorly understood.  
In this work, we propose that orthogonal glide symmetries can protect two distinct types of bulk anisotropic DPs. 
One is the ordinary anisotropic DP that is doubly degenerate only along one invariant axis, which is resulted from symmetry-protected accidental band crossing.
The other one is a symmetry-enforced DP fixed at a non-time-reversal-invariant point, which is doubly degenerate at three orthogonal directions. 
This unique topological phase  is exemplified by KSnSe$_{2}$ in space group 108. 
Our work not only unveils an unique symmetry protection mechanism but also provides the first material candidate for exploring multi-type Dirac fermions in noncentrosymmetric systems.  
\end{abstract}

\maketitle

{\it{Introduction.}} -- 
The classification and characterization of condensed phases of matter based on the topology of band structure has greatly enriched the contemporary understanding of semiconductors and insulators~\cite{RevModPhys.82.3045, RevModPhys.83.1057}.
Different topological states can posses the same symmetry but cannot be smoothly transformed from one to another, as they are distinguished by a global quantity like the number of holes on a sphere, based upon which the topology is defined. 
Time-reversal symmetry (${\cal T}$) and crystalline symmetry further extend the topological classification to symmetry-protected topological states. 
The pivotal role played by the crystalline symmetry even leads to the protection of bulk linear band degeneracy, i.e. a quantity distinct to the topological surface Dirac cone.
In these systems, the bulk-boundary correspondence yields the possible appearance of fermi arc. 
These quasiparticles mimic the fermionic particles in high-energy physics, such as the Dirac~\cite{PhysRevB.85.195320, PhysRevB.88.125427,zhongkaicd,Xu294,PhysRevLett.108.140405, PhysRevB.95.075135}, Weyl~\cite{PhysRevB.83.205101,PhysRevX.5.011029,Huang2015,Xu613,PhysRevX.5.031013,weyl-natphy} and Majorana fermions~\cite{Mourik1003,Nadj-Perge602}.  

In this work, we will focus on the four-fold linear band crossings, i.e. bulk DPs  in semimetals without inversion symmetry, whose protection imposes stronger symmetry requirement than for Weyl points. 
It is, thus, more difficult to realize. 
Crystalline-symmetry-protected bulk DPs have been extensively explored in systems with both ${\cal T}$ and spatial inversion symmetry (${\cal P}$). 
The joint operation of ${\cal T}$ and ${\cal P}$ leads to the Kramers degeneracy everywhere in the Brillouin Zone (BZ).
When any two of such bands cross, their crossing point can potentially be a DP under the presence of additional crystalline symmetries~\cite{1929PhyZ}.  
Such symmetries in centrosymmetric systems include rotation ($\hat{c}_{3}$, $\hat{c}_{4}$ and $\hat{c}_{6}$)~\cite{PhysRevLett.108.140405,Yang2014}, screw and glide mirror symmetries~\cite{PhysRevLett.115.126803, PhysRevB.93.155140}.
Compared to this, the symmetry protection mechanism for DPs in noncentrosymmetric system has far less been explored.
Only recently, it was shown that the nonsymmorphic operations can protect bulk DP in the absence of ${\cal P}$~\cite{Furusaki_Weyl_2017, PhysRevLett.121.106404, PhysRevB.99.201110}.
We further show in this work that multi-type stable bulk DPs can emerge in noncentrosymmetric systems under the protection of orthogonal glide symmetries and is further exemplified by a concrete material system in space group (${\cal SG}$)108. 
Our work not only unveils the coexistence of different bulk DPs but also provides the first material platform to explore multi-type Dirac fermions in noncentrosymmetric systems.
 
\begin{figure*}
\centering
\includegraphics[width=\linewidth]{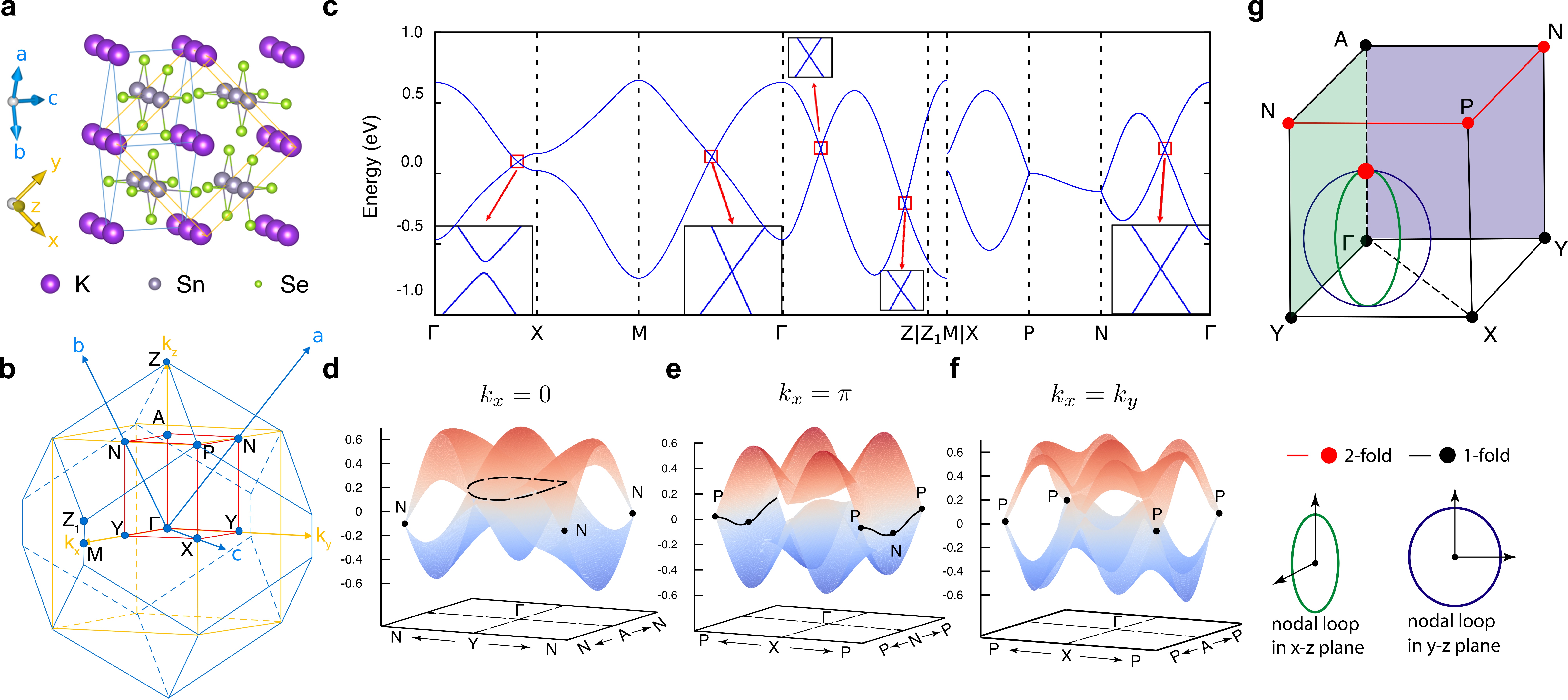}
\caption{(Color online) \textbf{Crystal, electronic structure of KSnSe$_2$ and symmetry-enriched nodal-loop in the absence of SOC}. (a) Primitive and conventional crystal cells. (b) The BZ of the primitive cell (denoted by the blue lines and axes) and conventional cell (denoted by the yellow lines and axes). (c) The electronic structure in the absence of SOC with zoom-in plots at each band touching points.  (d-f)  The symmetry enriched 2-fold nodal-point and nodal-loop at different planes with a schematic summary of band degeneracy of ${\cal SG} 108$ without SOC shown in (g). }
\label{Fig1}
\end{figure*}

 {\it{Symmetry analysis.}} -- 
The prototype material in ${\cal SG}$ 108 is KSnSe$_{2}$ with the unit cell shown in Fig.~\ref{Fig1}(a) and the BZ shown in Fig.~\ref{Fig1}(b). 
For the convenience of symmetry analysis, the conventional cell and the corresponding BZ are also displayed.
The band degeneracy will be only discussed in the BZ of the primitive cell.
From now on, we use $k_{x/y/z}$ to denote the momentum in cartesian coordinate and $k_{a/b/c}$ for it in terms of the reciprocal unit vectors of the primitive cell. 
A more detailed explanation of the relationship between conventional and primitive BZ can be found in the Supplemental Information.

The generators of ${\cal SG}$ 108 contain two groups of orthogonal glide symmetries given as follows:
\begin{eqnarray}
&&\begin{cases}
&\hat{g}_{x}: (x, y, z) \rightarrow (-x, y, z+\frac{1}{2})\otimes(-i\sigma_{x})\:,\\
&\hat{g}_{y}: (x, y, z) \rightarrow (x, -y, z+\frac{1}{2})\otimes(-i\sigma_{y})\:, 
\end{cases} \\
&&\begin{cases}
&\hat{g}_{xy}: (x, y, z) \rightarrow (-y, -x, z+\frac{1}{2})\otimes(-i\frac{\sqrt{2}}{2})(\sigma_{x}+\sigma_{y})\:,\\
&\hat{g}_{x\bar{y}}: (x, y, z) \rightarrow (y, x. z+\frac{1}{2})\otimes(-i\frac{\sqrt{2}}{2})(-\sigma_{x}+\sigma_{y})\:.
\end{cases}
\end{eqnarray}
The  Pauli matrices $\sigma_{x}$ and $\sigma_{y}$ denote the corresponding operations on spin. 
For spin-$1/2$ system, the half-lattice translation in $\hat{g}_{x}$ and $\hat{g}_{y}$ is known to protect Weyl node in systems with broken inversion symmetry, as the bands carrying different eigenvalues of the nonsymmorphic symmetry always appear in pair. 
Furthermore, they will have to switch their eigenvalues when the two Bloch bands evolve along a closed symmetry-invariant $k$-path, which leads to an unavoidable crossing point. 
The 2-fold crossing point is guaranteed to be stable owing to the different eigenvalues of the two bands. 
Moreover, the double operation of $\hat{g}_{x}$ or $\hat{g}_{y}$ yields a unit translation that makes the $k_{z}=0$ and $k_{z}=G_{z}/2$ planes distinct. 
Depending on the presence of spin-orbit coupling (SOC), one can have either a nodal-loop semimetal or a topological Dirac semimetal in ${\cal SG}$ 108, which we will discuss separately as follows. 

{\it 1. Nodal-loop semimetal without SOC.} -- 
In the absence of SOC, $\hat{g}_{x}$ and $\hat{g}_{y}$ no longer operate on the spin space and ${\cal T}^{2} = 1$. 
 At $k_{x/y}=0$ plane, ${\hat{g}_{x/y}}$ commutes with the Hamiltonian. 
 Each Bloch band carries definite eigenvalue $\lambda(\hat{g}_{x/y})=\pm e^{ik_{z}/2}$, which evolves from $\pm1$ at $k_{z}=0$ to $\pm  i$ at $k_{z}=G_{z}/2$. 
 Any two crossing bands with different $\lambda(\hat{g}_{x/y})$ cannot hybridize and will form a stable node. 
 This applies to every $k$-path of fixed $k_{y/x}$ with $k_{z}$ changing from 0 to $G_{z}/2$ at $k_{x/y}=0$ plane. 
 Consequently a stable nodal-loop forms. In Fig.~\ref{Fig1}(d) an example around the Fermi level at $k_{x}=0$ plane is shown as the dashed line.  
 Similar nodal-loop can be found at $k_{y}=0$ plane.
 However, we note that such stable nodal-loop appears accidentally.  
 It is not guaranteed that two bands will always cross at these two planes.  
 
 In contrast, $N$-point in Fig.~\ref{Fig1}(d) is always doubly degenerate. In fact, the entire $N-P$ line shown in Fig.~\ref{Fig1}(e) has a symmetry-enforced double degeneracy, which stems from the protection of ${\cal T}$ and $\hat{g}_{x/y}$.  
Under antiunitary transformation $\hat{\Theta}_{x}={\cal T}\hat{g}_{x}$,  $(k_{x},  {G}_{y}/2,  {G}_{z}/2)$ becomes $(k_{x}, - {G}_{y}/2, - {G}_{z}/2)$ which only differ by one primitive unit vector $ {G}_{b}$.
 $\hat{\Theta}_{x}$ is, thus, a symmetry operation leaves $k$-point along $(k_{x},  {G}_{y}/2,  {G}_{z}/2)$ invariant. 
Similarly, $\hat{\Theta}_{y}={\cal T}\hat{g}_{y}$ can be defined as the symmetry operation for $( {G}_{x}/2, k_{y},  {G}_{z}/2)$ line.
It is easy to show that $[\hat{\Theta}_{x/y}]^{2}=e^{ik_{z}}=-1$ at $k_{z}= {G_{z}}/2$, which indicates a Kramers pair of states $| \psi\rangle$ and $\hat{\Theta}_{x/y}| \psi\rangle$.
Such degeneracy is symmetry-enforced, which is always present and stable. 
While, although $A-N$ is also at $k_{z}= {G_{z}}/2$ plane, it is not $\hat{\Theta}_{x}$ invariant as  $(k_{x}, 0,  {G}_{z}/2)$ is not equivalent to  $(k_{x}, 0, - {G}_{z}/2)$. 
The same argument applies to $\hat{\Theta}_{y}$ on $A-N$ as well. 
At $k_{z}=0$ plane, either due to the lack of $\hat{\Theta}_{x/y}$ invariance or $[\hat{\Theta}_{x/y}]^{2}=1$, one will always have non-degenerate state. 
We summarize the symmetry-protected band degeneracy of ${\cal SG}$ 108  under the absence of SOC in Fig.~\ref{Fig1}(g). 
 
 \begin{figure*}
\centering
\includegraphics[width=\linewidth]{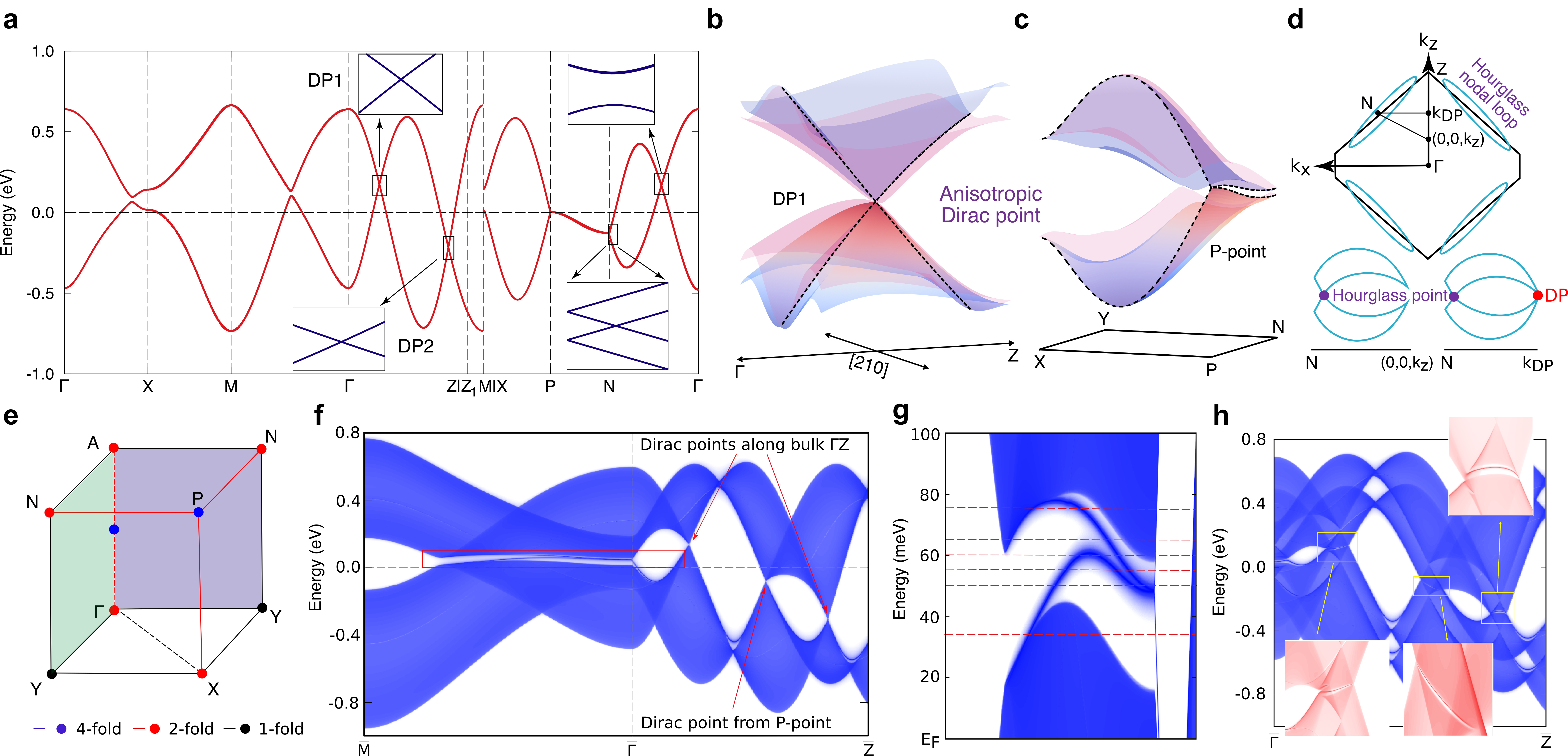}
\caption{(Color online) \textbf{Electronic structure and the topological surface states in the presence of SOC} (a) Bulk electronic structure around the Fermi level. (b) The anisotropic bulk DP1 along $\Gamma-Z$. (c) Same as (b), but at $P$-point.  (d) The emergence of the hourglass loop between N and $\Gamma-Z$ line. To a better visualization, we have significantly enlarge the hourglass loop in this plot. (e) A schematic summary of band degeneracy protected by ${\cal SG}$-108 with SOC. (f, g) The topological surface states with a zoom-in plot around $\bar{\Gamma}$. (h) Topological surface states calculated with small Zeeman field. Each DP splits into two Weyl points with notable Fermi arc connecting in between, which characterizes the topological nature of these DPs. }
\label{Fig2}
\end{figure*}

 {\it 2. Topological Dirac semimetal with SOC.} --
 Without SOC, the two orthogonal glide symmetry $\hat{g}_{x}$ and $\hat{g}_{y}$ do not jointly protect any additional degeneracy.
While, when SOC is present, ${\cal T}^{2}=-1$ and the double group operation will lead to additional symmetry-enforced band degeneracies, including the emergence of a symmetry-protected DP along $\Gamma-A$ and a symmetry-enforced DP at $P$-point which is not a time-reversal invariant momentum (TRIM). 
They form two different types of DPs distinct to each other as follows:
\begin{enumerate}
\item The DP along $\Gamma-Z$ can be shifted along this line without violating any symmetry, i.e. its location is not fixed. While the other one is fixed at $P$-point. 
\item As for the one at $\Gamma-Z$, the bands are doubly degenerate only along $\Gamma-Z$. 
While, at $P$-point, the double band degeneracy is kept along three orthogonal axises. 
\end{enumerate}
 
 {\it $\Gamma$-Z [(0, 0, $k_{z}$)]}. -- Due to the additional operation on spin, along $\Gamma-Z$ any Bloch state carries a definite eigenvalue of $\hat{g}_{x}$ as $\pm ie^{ik_{z}/2}$, i.e. $\hat{g}_{x}| \psi^{\pm}\rangle=\pm ie^{ik_{z}/2}| \psi^{\pm}\rangle$ where $\pm$ denotes the states with eigenvalue of positive or negative sign. 
Owing to the anti-commutation $\{\hat{g}_{x}, \hat{g}_{y}\}=0$, $\hat{g}_{y}| \psi^{\pm}\rangle$ becomes orthogonal to $| \psi^{\pm}\rangle$, i.e. $\hat{g}_{x}(\hat{g}_{y}| \psi^{\pm}\rangle) = -\hat{g}_{y}\hat{g}_{x}| \psi^{\pm}\rangle = \mp ie^{ik_{z}/2}\hat{g}_{y}| \psi^{\pm}\rangle$. 
As a result, any state $| \psi\rangle$ along $\Gamma-Z$ doubly degenerates with state $\hat{g}_{y}| \psi\rangle$, and they carry opposite $\hat{g}_{x}$ eigenvalues. 
If any two such doubly degenerate bands cross along $\Gamma-Z$, it will be possible to generate a DP under $\hat{c}_{4}$ rotational symmetry.
In this sense it is similar to the ${\cal TP}$-invariant system with rotational symmetry, however, here the doubly degenerate bands are induced by two orthogonal glide operations instead of the normal ${\cal TP}$ operation.  
Nevertheless, it is the same as ${\cal TP}$-symmetric systems that the location of the DP along this line is not predictable by symmetry, i.e. it can be anywhere depending on the detailed chemical environment. 


\begin{figure*}[t]
\centering
\includegraphics[width=\linewidth]{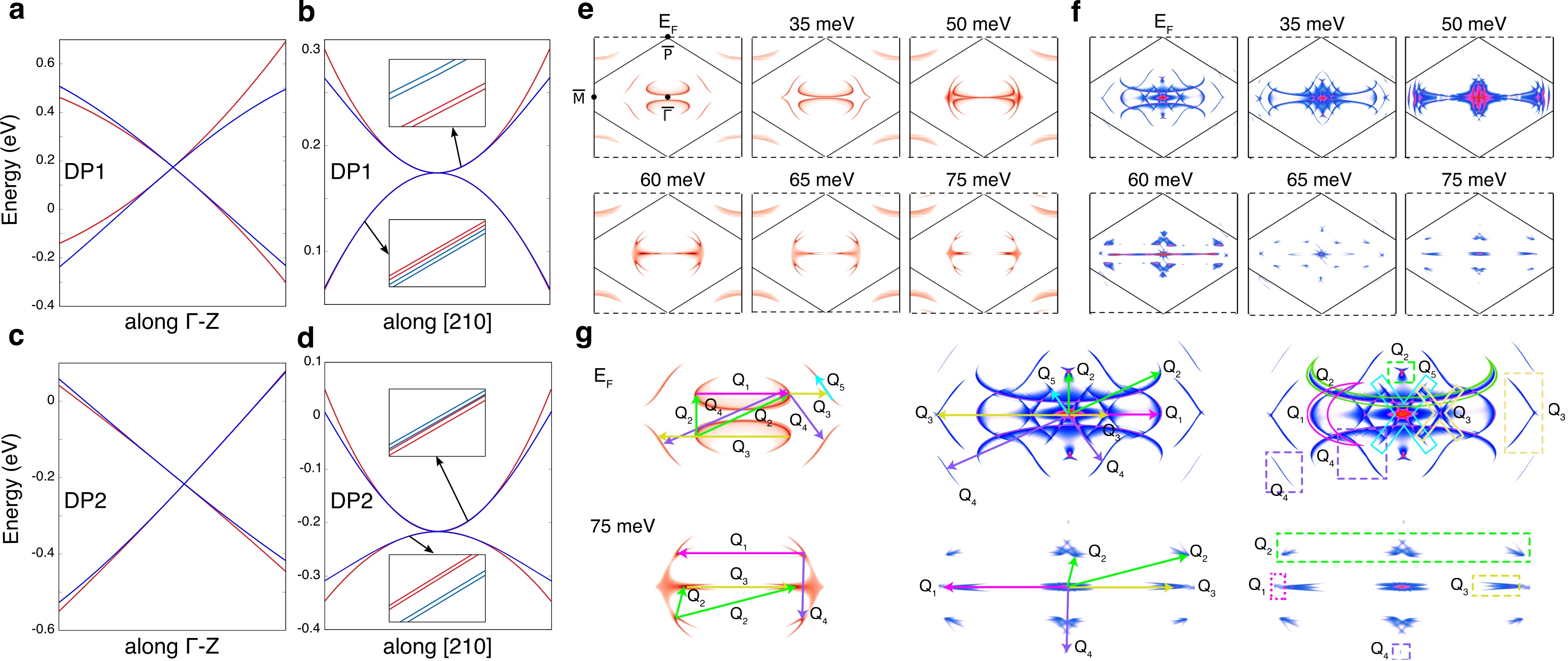}
\caption{(Color online) \textbf{$k\cdot p$ model, iso-energy cuts and the QPI patterns.} (a-d) The band structure near the two DPs along in-plane and out-of-plane directions calculated from the $k\cdot p$ Hamiltonian (red lines) with fitted parameters obtained in DFT (blue lines). Every band is doubly degenerate along $\Gamma-Z$ and singly degenerate along [210]. (e) The iso-energy cuts  of the topological surface states and (f) the corresponding QPI patterns. (g) Magnified view of the iso-energy plot of the topological surface states and QPI patterns on the Fermi level (the first row) and on $+75$ meV (the second row). The $Q$ vectors indicate the scattering wave vectors, and the rectangles denote the area in which the features are dominated by the vectors with the same colors. }
\label{Fig3}
\end{figure*}

{\it P-point}. -- Due to the lack of time reversal invariance, $P=( {G_{a}}/4,  {G_{b}}/4,  {G_{c}}/4)$ is not expected to host high band degeneracy.  
However, in ${\cal SG}$ 108 the presence of the orthogonal glide symmetries make this point a DP. 
To the best of our knowledge, symmetry-enforced DP fixed at non-TRIM has not been discussed before, which is completely protected by two groups of orthogonal glide symmetries. 

The little group at $P$-point consists of $\hat{g}_{xy}$, $\hat{g}_{x\bar{y}}$, $\hat{\Theta}_{x}$, $\hat{\Theta}_{y}$ and $\hat{c}_{2}$.
The allowed eigenvalues are $\lambda({\hat{g}_{xy}})=\lambda(\hat{g}_{x\bar{y}})=\pm1$ and $\lambda({\hat{\Theta}_{x}})=\lambda(\hat{\Theta}_{y})=\lambda(\hat{c}_{2})=\pm i$. 
If one assumes a state $|\psi_{1}\rangle$ with $\lambda(\hat{g}_{xy}) = 1$ and $\lambda(\hat{c}_{2})=i$, another three states orthogonal to $|\psi_{1}\rangle$ can be constructed as $|\psi_{2}\rangle=\hat{g}_{x\bar{y}}|\psi_{1}\rangle$, $|\psi_{3}\rangle=(\hat{\Theta}_{x}+\hat{\Theta}_{y})|\psi_{1}\rangle$ and $|
\psi_{4}\rangle=(\hat{\Theta}_{x}-\hat{\Theta}_{y})|\psi_{1}\rangle$ from the anticommutation relation $\{\hat{g}_{xy}, \hat{g}_{x\bar{y}}\}=0$ and the antiunitary property $\hat{\Theta}_{x}^{2}=\hat{\Theta}_{y}^{2}=-1$. 
In fact, under such construction $|\psi_{2}\rangle$, $|\psi_{3}\rangle$ and $|\psi_{4}\rangle$ are also mutually orthogonal to each other. 
One can easily prove that they carry different eigenvalues of $\hat{g}_{xy}$ and $\hat{c}_{2}$. 
Under $\hat{c}_{2}$, $|\psi_{2}\rangle$ carries eigenvalue $-i$ but $i$ for $|\psi_{3}\rangle$ and $|\psi_{4}\rangle$ due to $\{\hat{g}_{x\bar{y}}, \hat{c}_{2}\}=\{\hat{\Theta}_{x}, \hat{c}_{2}\}=\{\hat{\Theta}_{y}, \hat{c}_{2}\}=0$.
Finally, $|\psi_{3}\rangle$ and $|\psi_{4}\rangle$ take different eigenvalues of $\hat{g}_{xy}$ as 1 and -1 due to the cross commutation-relation $ \hat{g}_{xy}\hat{\Theta}_{x}=\hat{\Theta}_{y}\hat{g}_{xy}$ and $\hat{g}_{xy}\hat{\Theta}_{y}=\hat{\Theta}_{x}\hat{g}_{xy}$, see the Supplementary Information for the proof. 

Similarly, one can prove that states on {\it X-P} and {\it N-P} are two-fold degenerate guaranteed by $\hat{g}_{xy} (\hat{g}_{x\bar{y}})$ and $\hat{\Theta}_{x}(\hat{\Theta}_{y})$. 
$P$-point, as the crossing point of $X-P$ and $N-P$ is, thus, protected by two groups of orthogonal glide symmetries.
Along the three orthogonal $X-P$ and $N-P$ lines,  the double band degeneracy is protected, which defines a new type of anisotropic DP distinct to the one along $\Gamma-Z$ whose double degeneracy is only present along $\Gamma-Z$, see Fig.~\ref{Fig2}(b) and (c) for the comparison. 
For better visualization, we artificially enlarged the inversion asymmetry by hand to increase the band splitting. 
Similar as in other nonsymmorphic systems, we also find a hourglass type nodal-loop in KSnSe$_{2}$ shown in Fig.~\ref{Fig2}(d) around $N$-point. 
In KSnSe$_{2}$ the energy splitting for bands between $N$ and $(0, 0, k_{z})$ is too small to have significance and the hourglass nodal-loop is barely observable in experiment.   
A summary of band degeneracy is illustrated in Fig.~\ref{Fig2} (e). 

{\it{Topological character}}.-- 
Fig.~\ref{Fig1}(c) and Fig.~\ref{Fig2}(a) display the top two valence bands and the bottom two conduction bands of KSnSe$_{2}$ obtained in the calculations without/with SOC, respectively. 
These bands isolate from all other bands in energy.
All exciting physics can be demonstrated in this low-energy sector and has been partially discussed in the previous section. 
The nodal loop at $k_{z}=0$ plane is removed by SOC, resulting in gapped electronic structure everywhere in this plane with a band inversion at $\Gamma$.  
As displayed in Fig.~\ref{Fig2}(f) and the zoom-in plot in Fig.~\ref{Fig2}(g), topological surface states appear inside this gap on the $(001)$ surface.
Along $\Gamma$-$Z$, each band is doubly degenerate.
Two bands accidentally invert and cross twice with stable crossing points even in the presence of SOC. 
As explained before, these accidental crossings are protected jointly by the two orthogonal glide operations $\hat{g}_{x/y}$ and $\hat{c}_{4}$ rotation.

To get more insight of this anisotropic DP, we further derive a low-energy effective Hamiltonian.
The symmetry operations considered here include $\hat{c}_4$, $\hat{g}_{x/y}$, and ${\cal T}$. 
As known from density functional theory (DFT) calculations, the states near the Fermi level are mainly $|\Gamma_{6}, \pm3/2\rangle$, $|\Gamma_{7}, \pm1/2\rangle$.
Based upon this basis, we obtain the effective $k\cdot p$ Hamiltonian for the DP along $\Gamma-Z$:
\begin{equation}
H(\vec{k})=
\left(
  \begin{matrix}
   M_+(\vec{k}) & B_+(\vec{k}) &0  & 0 \\
   \dagger & M_-(\vec{k}) &0 &0 \\
    0 & 0 & M_+(\vec{k}) & B_-(\vec{k})\\
    0 &0 & \dagger&M_-(\vec{k})
      \end{matrix}
  \right),
\end{equation}
where $M_{\pm}(\vec{k})=C \pm M_0+(M_1 \mp M_3)(k_x^2 + k_y^2)+(M_2 \mp M_4)k_z^2$ and $B_{\pm}(\vec{k})=B_1k_xk_y \mp iB_2(k_x^2 - k_y^2) - iB_3k_xk_yk_z \pm B_4k_z(k_x^2-k_y^2)$.
Coefficients $C, M_0, M_1, M_2, M_3, M_4, B_1, B_2, B_3, B_4$ are parameters which can be easily obtained by fitting the model to the DFT calculations, see the Supplementary Information for more details.  
This $k\cdot p$ model successfully captures the characteristic anisotropy of the DPs as shown in Fig.~\ref{Fig3} (a-d).
Interestingly, unlike the $k\cdot p$ models for most of the other topological systems, the cubic terms must be included to break inversion symmetry. 
For the basis we have ${\cal P}=-\sigma_0\otimes\tau_0$, and only the two cubic terms are anticommute with it. As one can expect, the hamiltonian up to quadratic term gives double-degeneracy in the whole BZ as it commutes with both ${\cal P}$ and ${\cal T}$. 
Including the cubic term splits each double-degenerate bands except along $\hat{c}_4$ axis. 
Hence the coefficients of the cubic terms, $B_3, B_4$, are "the degrees of inversion asymmetry".

Concerning the topological nature of these DPs, unlike Weyl point which carries definite chiral charge, DP is a composition of two Weyl points and contains no net chiral charge.
And DPs are not promised to have arc states in between. Thus, it is often difficult to characterize their topological nature. 
Here, we propose to examine the presence of Fermi arc after splitting a DP into two Weyl points with small Zeeman field, by which we find that both types of DPs observed in this space group are topologically nontrivial.
Similar as Fig.~\ref{Fig2}(f), Fig.~\ref{Fig2}(h) displays the surface states along $\bar{\Gamma}-\bar{Z}$ but with Zeeman splitting. 
Each DP splits to two Weyl points with notable fermi arc in between as shown by the zoom-in plot at each split DP. 

Fig.~\ref{Fig3} (e) shows the iso-energy plot and the evolution of the topological surface states with energy levels indicated by the red dashed-lines in Fig.~\ref{Fig2} (g). 
Here the bulk contribution has been subtracted with only the surface contribution left. 
The specific arc-shape surface states yield the novel quasiparticle interference (QPI) patterns~\cite{qpi1,qpi2} shown in Fig.~\ref{Fig3}(f).
We calculated the QPI from the joint density of states (JDOS) $J(\vec{q})=\sum_{\vec{k}}{\rho(\vec{k})\rho(\vec{k}+\vec{q})}$, where $\rho(\vec{k})$ is the density of states at momentum vector $\vec{k}$ on the weighted iso-energy contours and $\vec{q}$ is the transfer momentum.
Remarkably, despite the fact that $\rho(\vec{k})$ contains both bulk and surface states, all wave vectors in the QPI patterns can be understood solely from the topological nontrivial surface states. 
As illustrated in Fig.~\ref{Fig3} (g) at the Fermi level and at $E=75$ meV, via the length and orientation of every two topological surface states we can nicely interpret all the strongly weighted JDOS patterns, which are ideal to be observed by scanning tunneling spectroscopy.

{\it Conclusions.} -- Through symmetry analysis, we have identified a new mechanism protecting bulk DPs in a noncentrosymmetric system of ${\cal SG}$-108. 
The two orthogonal glide symmetries, together with $\hat{c}_{4}$ rotation and time-reversal symmetries promote two distinct types of DP, i.e. the symmetry-protected DP along $\Gamma-Z$ and symmetry-enforced DP at non-TRIM {\it P}-point. 
Both types of DP are shown to be topologically nontrivial through the application of Zeeman field in surface states calculation. 
The appearance of additional arc states connecting the two Weyl points split from a DP provides a simple way to characterize the topology of a DP. 
The proposed material candidate KSnSe$_{2}$ nicely demonstrates all the features with experimentally detectable QPI pattern and topological surface states.
It provides the first material example to realize this unique symmetry protection mechanism and multi-type bulk DPs in noncentrosymmetric systems.

 {\it Acknowledgements} -- 
We would like to thank the useful discussion with S.Y. Yang on the $k\cdot p$ model. 
This work was supported by the National Key RD Program of China (2017YFE0131300).
G. Li would like to thank the financial support from the starting grant of ShanghaiTech University and the Program for Professor of Special Appointment (Shanghai Eastern Scholar).
Calculations were carried out at the HPC Platform of ShanghaiTech University Library and Information Services, as well as School of Physical Science and Technology.




\bibliographystyle{apsrev4-1}
\bibliography{ref}  

\end{document}